# Containers Placement and Migration on Cloud System

Oussama Smimite
LabSIV, Department of Computer Science
Faculty of Science, Ibn Zohr University
BP 8106, 80000 Agadir, Morocco

Karim Afdel
LabSIV, Department of Computer Science
Faculty of Science, Ibn Zohr University
BP 8106, 80000 Agadir, Morocco

## ABSTRACT

Currently, many businesses are using cloud computing to obtain an entire IT infrastructure remotely while delegating its management to a third party. The provider of this architecture ensures the operation and maintenance of the services while offering management capabilities via web consoles.These providers offer solutions based on bare metal or virtualization platforms (mainly virtual machines). Recently, a new type of virtualization-based on containerization technology has emerged. Containers can be deployed on bare metal servers or in virtual machines. The migration of virtual machines (VMs) and Containers in Dynamic Resource Management (DRM) is a crucial factor in minimizing the operating costs of data centers by reducing their energy consumption and subsequently limiting their impact on climate change.In this article, live migration for both types of virtualization will be studied. for that, container placement and migration algorithms are proposed, which takes into account the QoS requirements of different users in order to minimize energy consumption. Thus, a dynamic approach is suggested based on a threshold of RAM usage for host and virtual machines in the data center to avoid unnecessary power consumption. In this paper, the proposed work is compared with VM/Container placement and migration methods, the results of the experiment indicate that using container migration instead of VMs demonstrates a reduction in power consumption, and also reduces the migration time which impacts QoS and reduces SLA violation.

## 1. INTRODUCTION

Cloud computing has multiple advantages for IT in terms of portability, upgradability, scalability, high availability, resource sharing and most of all, cost efficiency [29]. For these reasons, operators and software providers are switching to cloud computing for their infrastructure to gain more flexibility while optimizing their costs. On the ecological level, according to Energy Information Administration, (World Energy Projections Plus), The IT industry accounts for 2% of global CO2 emissions and IT related CO2 emissions are rising 5.5 times faster than global CO2 emission levels. Moreover, the energy consumption of data centers is higher than in other office buildings. With this in mind, service providers in a cloud environment should take steps to avoid performance degradation and high power consumption [28].

For cloud data centers, virtualization technologies such as Xen , Hyper V and VMware are widely used due to their ease of use, cost efficiency, flexibility of resource management, scalability and the simplicity of enabling the high availability [27]. More precisely, virtualization technologies offer the possibility of fine-tuning the resources allocation by associating processors, RAM, disk space and network bandwidth to a specific Virtual Machine [30, 6]. This approach has allowed the development of solutions such as Software as a Service (SaaS) and Platform as a Service (PaaS), on top of the usual Infrastructure as a Service (IaaS), where services providers can quickly make available virtual machines with the required resources to their customers almost in no time, and not burden them with the pain of infrastructure management.

In some scenarios, for servers maintenance period, load balancing across nodes, fault tolerance and energy management on the datacenter, administrators are required to perform migration from one node to another. Accordingly, migration has become of the most important aspects of virtualization and even a powerful selling argument for many cloud services providers as it translates into their Service Level Agreement (SLA).

Operating Systems (OS) instances migration across different physical hosts is a useful feature for Cloud Data Centers administrators. By performing migration while OS continues to operate, service providers can guarantee high performances with minimal application downtime. It can actually be so small that a virtual machine user does not even notice it. However, as small as the downtime might be, it can cause some serious issues for time-sensitive applications such as online gaming, audio/video streaming, banking or any other time-critical web applications. Therefore, this degradation is simply not acceptable for these service and cause harm to the quality of service and service level agreement offered by the provider. Thus, it is vital to understand the migration procedure and techniques in order to integrate them efficiently and reduce their impact on the quality of service [29].

In this paper, the live migration mechanisms of VMs and containers are presented, after that, a strategy of allocation and migration of containers is proposed in order to reduce the number of VMs and physical hosts. The rest of this document is grouped as follows. In Section 2, community related works are discussed. Section 3 presents virtualization and containerization technologies as well as types of live migration. Section 4 presents a detailed proposed system model and proposed policy parameters for performing the live migration. The analysis of experimental results and evaluation are presented in section 5. In the last section, conclusions and future work are discussed.

## 2. RELATED WORKS

The development of new technologies is leading to an explosion in computer data storage needs and an exponential increase in server power consumption. In order to reduce power consumption, it is necessary to improve the use of these servers by putting a maximum of virtual machine / container in these host machines while respecting the SLA contracts. This optimization is generally carried out by load balancing and migration algorithms. Several works have been carried out in this field. [16, 31, 8, 26, 22].

In the research paper "Comparing Live Migration between Linux Containers and Kernel Virtual Machine", the authors have Compared Live Migration between LXD and KVM on the aspects of CPU usage, disk utilization, downtime and total time migration. Based on the results of the study, it can be stated that KVM has relatively better CPU performance when compared to KVM. Whereas LXC has relatively better performance than KVM when downtime, total migration time, and disk utilization are considered [18].

Kamran et al. have presented algorithms and models for virtual machine placement and migration. The simulation results demonstrate a reduction in power consumption, reduce the number of VM migrations and decrease SLA violations when the QoS-aware MM algorithm is used. [23].

Sareh Fotuhi et al. have compared host selection algorithms for container and VM considering three metrics including energy consumption, average number of migrations and SLA violations. Results show in a containerized cloud environment where container consolidation is available, migrating containers is more energy efficient than consolidation virtual machines with minimal SLA violations [24].

Min Su Chae et al. have compared Docker containers with virtual machines and measured the performance of the two technologies. The results showed that Docker requires minimum resources to run the application in Containers, and uses hardware resources such as RAM, CPU, HDD more efficiently and faster than KVM. [11].

Bilal Ahmad et al. have compared three different power consumption approaches," non-power aware, dynamic voltage frequency scaling and static threshold virtual machine consolidation technique", to assess the most optimal technique in terms of energy consumption and compliance with service level agreements. The result shows that dynamic voltage frequency scaling saves more energy as compared to other approaches. It has fewer SLA violations which are important for maintaining QoS.In addition, the use of the DVFS technique in simulation environments such as CloudSim [9] shows that data centers can consume less energy, which provides more economical solutions. [4].

All the above-mentioned works mainly use the migration of virtual machines or bare metal containers but they do not deal with hybrid virtualization where the placement strategy plays an important role.

## 3. VIRTUALISATION

Virtualization is a technology for creating and executing one or more virtual representations of a host or its different resources on the same physical machine. The virtual machines are run on a hardware abstraction layer. Thus, several operating systems can be run simultaneously on the same server. The resources and capabilities of the server are distributed among the different instances. This makes it possible to use servers more efficiently because their capabilities are fully exploited. In addition, virtualization allows programs to run in isolation to perform tests safely without compromising other virtual machines running on the same server .

### 3.1 Virtual Machine (VM)

In Virtual machines, the applications running in different VMs are isolated from each other make it easy to move applications by transferring images from one server to another. This virtualization offers some security benefits and can reduce management complexity and thus provide a consistent environment for the entire infrastructure when all applications are on the same type of VM, even if Host servers are not homogeneous. Still, virtual machines have some disadvantages such as:

—Server resources may be underutilized;
—In most cases, VMs cannot access physical hardware directly;
—Virtual machines typically do not work as well as physical host servers because of the abstraction layer between the application and the hardware [1].

### 3.2 Container deployed in bare-metal

Running containers on bare metal hosts without an OS has many advantages, such as high performance for applications, because there is no hardware emulation layer separating them from a host server. Furthermore, it makes it possible to deploy applications in environments that can easily switch from the host server to another, and finally the containerization allows application isolation. If containers may not offer the same level of isolation as VMs, they still prevent applications from interacting with each other and setting strict limits on the privileges and accessibility of resources associated with each container. But there are major obstacles that prevent the direct deployment of containers on bare-metal host's server, such as the problem of updating physical servers. In fact, to replace a bare metal server, you must recreate the container environment from scratch while using virtualization makes it easy to migrate VMs to a new server. Another problem is that the containers depend on the type of the operating system, for example Linux containers run in Linux hosts and Windows containers run on Windows hosts. Virtualization platforms also make it possible to create a virtual machine snapshot, which gives the possibility of rollback, which is not possible for bare-metal servers that do not provide rollback functions [1].

### 3.3 Container deployed in Virtual machines (Hybrid virtualization)

Due to the fact that containers on the same physical host share the operating system kernel, a container security violation might compromise and affect other applications that share the same physical host. More precisely, the isolation in containers is a software isolation nature, unlike VM where it is of a hardware isolation type, the security aspects still a bit of a challenge. Because of these aspects, service providers do not offer the same migration capabilities, as it is the case for virtual machines. To remedy these problems, nesting a container in a virtual machine allows administrators to manage each container separately using a simple model: one container per VM. This allows using the existing virtualization management software and keeping the existing process. In addition, because each container relies on a VM as an additional layer of abstraction, administrators can avoid the security issues of multiple containers sharing the same OS kernel and maintain administrative consistency. This Hybrid architecture [25, 10, 7, 14] raises questions about the efficiency of combining containers and VMs, and how to optimize the capacity usage of the hardware infrastructure. In other words, if virtual machines and containers can coexist, and if so, can the container services interact with VM services properly. In fact, the experience described below shows that VM machines and containers can coexist and that container services can interact with the services of the VM virtual machine without any particular difficulty [3]. For all this working on a hybrid system that allows to benefit from VMS and containers is recommended.

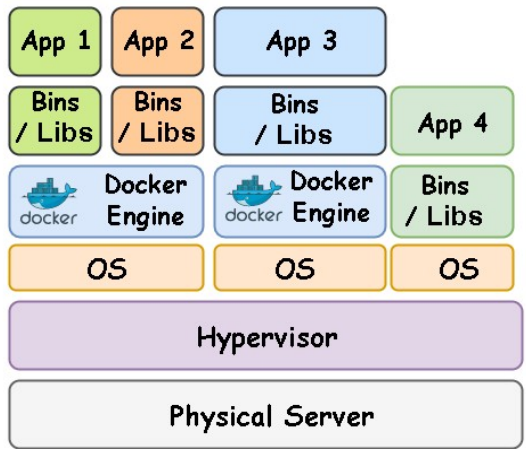

Fig. 1. Nesting of containers integrating applications in VMs.

### 3.4 Live Migration

3.4.1 *Migration of Virtual Machine (VM).* Migration allows a clear isolation between Hardware and Software and optimizes portability, fault tolerance, load balancing and low level system maintenance [20]. VM migration can be classified into two categories [17]:

—Cold migration, when an administrator is planning to upgrade a piece of hardware or perform a maintenance task on a cluster,the customer must be notified in advance about the maintenance schedule and its duration. Then, the hardware is turned off and back on only when the system and data are successfully restored. This approach is bothersome as it is not compliant with High Availability as the downtime is generally important.
—Live migration, it based on the process of transferring a virtual machine or a container while running from a physical node to another with a minimal interruption time. The migration of the physical host can be achieved by moving the state of processor, network and storage toward a new one and even the content of the RAM.

The process of transporting a virtual machine from a physical server A to a new one B consists of five steps [17, 12, 21]:

(1) Reservation: Making sure that the destination server (B) has enough resources to host the VM to migrate and allocate them.
(2) Iterative pre-copy: Transferring data from memory pages of server (A) toward server (B). This action is then repeated only for the update pages (Dirtied) during the the previous transfer.
(3) Stop-and-Copy: Halt the Virtual Machine on the source server (A) then copy the remaining data to destination server (B).
(4) Engaging: Send a notification message from the server (B) to (A) to inform about the end of migration. The VM instance is deleted from the source sever (A).

(5) Activation: Enabling the VM that has been successfully migrated to server (B).

3.4.2 *Container Migration.* Containers technologies [18, 15, 19, 13] are generally very interesting due to there ability to simplify the portability of cloud applications: it allows the execution of an application across multiple platforms without worrying about low level dependencies or OS compatibility. This explains their attractiveness for multiple businesses as they offer unprecedented flexibility for application deployment.

Unlike hypervisor-based virtualization, container-based virtualization is performed at a low-level system and does not aim to emulate a full hardware environment. It relies on the Linux Kernel capabilities to isolate applications. With this level of virtualization, multiple isolated Linux systems (containerized) can run on a control host while sharing a single instance of the OS kernel.(Figure 2).

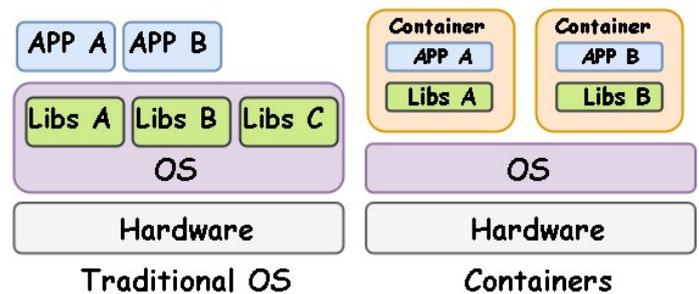

Fig. 2. Difference between traditional OS and Containers.

Each container has its own processes and own network. The isolation is achieved thanks to the name spaces. In fact, the processes of a container have unique identifier within their name space and cannot interact with those of another name space directly. A container can be seen as a set of processes (binaries, associated libraries, config files) that has its own lifecycle and dependencies and separated from other containers. Containerization is done today using tools like LXC (LinuX Containers) [15], Docker [19] or OpenVZ . Container migration is relatively a new technology on the market. In general, containerization is getting a lot of attractiveness from many businesses as a result of various benefits: almost no downtime maintenance, low effort for configuration and deployment, flexibility and the simplicity to achieve high availability. Unlike virtualization, containerization allows virtual instances (containers) to share a single host operating system, including binary files, libraries, or drivers. Thus, it allows a server to potentially host many more containers than virtual machines. The difference in hosting can be considerable, so a server can accommodate 10 to 100 times more container instances than virtual machine instances. Because an application container does not depend on the OS, it is significantly smaller, easier to migrate or download, faster to backup or restore. Moreover, by isolating software packages from each other in containers, it ensures a higher safety of sensitive applications [15, 19, 13].

Container migration is relatively complex, because containers are rather processes belonging to a namespace. Currently, container migration is based on the CRIU tool (Checkpoint/Restore In User Space). CRIU performs this by freezing all process states on the Host machine and then restores them in the destination machine according to the sequence diagram of Figure 3

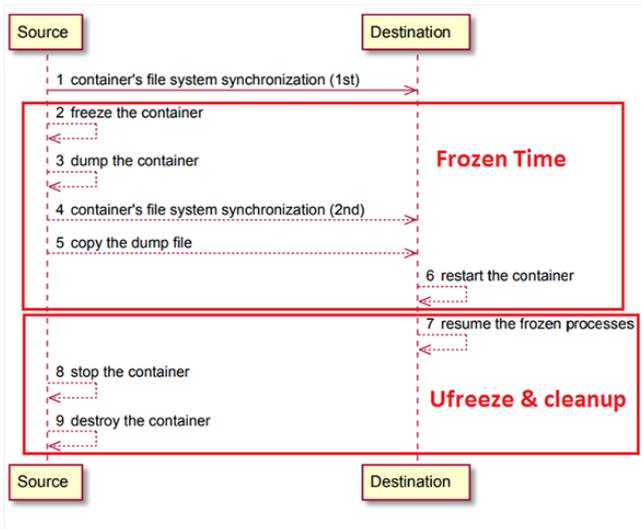

Fig. 3. The sequence diagram showing the Live Migration of the container.

## 4. PROPOSED STRATEGY

The key part of the system is the Broker Manger, the latter acting as an intermediary between service providers and users. A Cloud Broker can also negotiate SLA contracts with cloud providers instead of the customer. after receiving the user's request the broker initiates the processing and submits those requests to the other modules. It grants an interface to handle multiple clouds and resources sharing. Figure 4 illustrates the proposed system architecture .

### 4.1 Proposed System Architecture

Figure 4 shows the architecture used, which is composed of the following modules:

4.1.1 *The Brokers.* The broker is the main module for ensuring communication between service providers and users, and for providing services based on users' quality of service requirements, determine SLA requirements and arrange services on the cloud. This module is also in charge of analyzing user QoS service specification for the processor, RAM, and instance bandwidth performance.after that based on information delivered by the monitor module deciding even if to approve or reject the request.

4.1.2 *The Global Resource Manager.* The main role of this module is to analyze the requests submitted by users and check the QoS requirements before deciding to approve or deny the request. also, he decides for Requests have deadlines, whether he has enough time and credit to complete the work on time or not, if the work is finished on time, it is accepted if not rejected. To ensure that no SLA infringement appear the module request updated information about the VM and Containers Monitor to make resource allocation decisions efficiently. Then, it decides the right resources for the allocation of the Containers.

4.1.3 *The Monitor of containers and VM .* The container and VM monitor is responsible for monitoring containers, virtual machines, and servers. The main task of this module is to manage the state of resources, RAM usage, processor usage, SLA violation, and power consumption. Once an anomaly in the use of resources is triggered,The monitor sends a notification to the Resource Allocator SLA to take appropriate action.

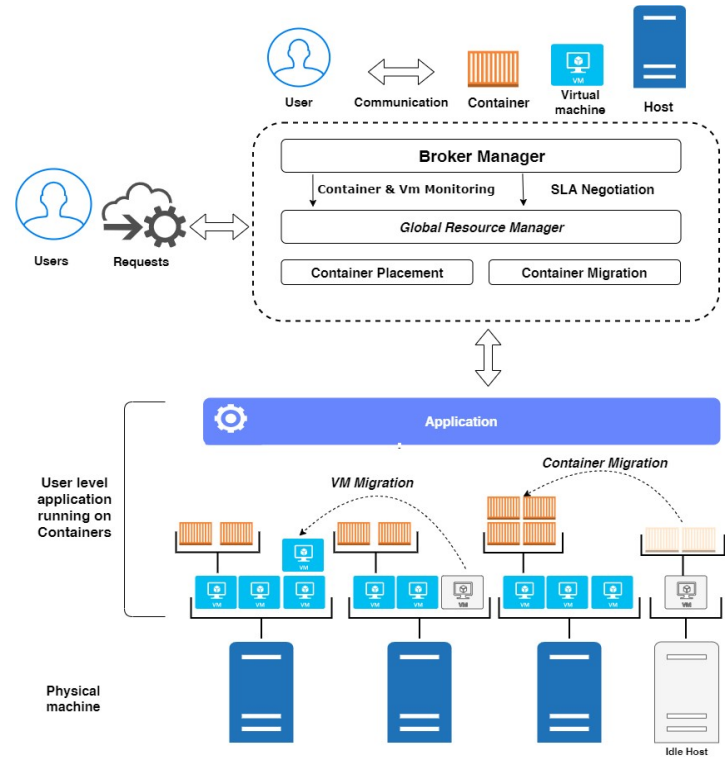

Fig. 4. Proposed system architecture.

4.1.4 *The Physical machines (PM).* The physical machine is also known as the "bare-metal server", which is the hardware support for creating and hosting virtual machines. depending on its capacity, the MP can host several VMs or Container.

4.1.5 *The Virtual machines (VM).* VM is a software used as a simulation of the physical machine, It is possible to run several virtual machines at the same time on the same physical machine. For servers, the various operating systems work side by side, with a software component called a hypervisor to manage them. VM migration is the operation of moving VMs from one PM to another. Virtual machine migration is used to improve performance, hardware usage, and energy usage.

4.1.6 *Container-based virtualization (CNT) .* Container-based virtualization is the virtualization layer runs within the operating system. Containers are thus exceptionally light and take just seconds to start, versus minutes for a VM. Containers speed, agility, and portability make them yet another tool to help streamline software development. Process of transferring Containers from one PM to another, or from VM to another called Container migration.

(1) Cloud Broker receives the user's request with QoS requirements, like response time, deadline or budget or, after that the broker manager layer uses scheduling mechanisms and admission control to accept or refuse this request.
(2) If the request is approved, the two parties sign a formal agreement (SLA) in order to ensure QoS requirements such as availability or response time.
(3) The cloudlets are allocated to the available Container.
(4) The result is submitted to the user once the cloudlet is successfully executed.

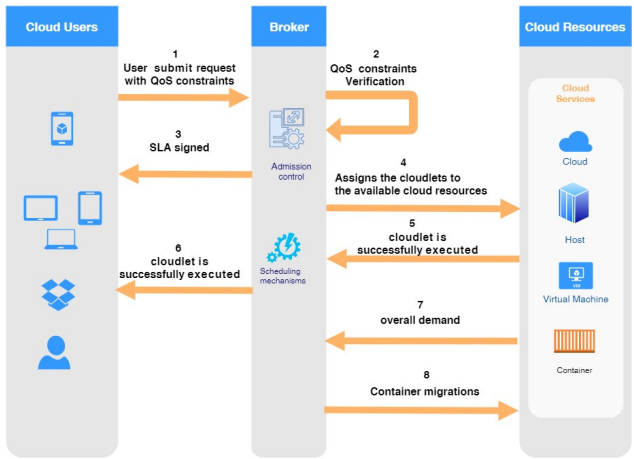

Fig. 5. The sequence diagram showing the Proposed strategy .

(5) If the total RAM demand exceeds the available RAM, the cloudlets are adjusted or allocated to a different resource This can cause unexpected delays, which affects the QoS requirements.
(6) To deal with this issue, the load balancing method is used by migrating Container from an overloaded VM to another appropriate VM.
(7) To save energy, all virtual machines or containers are migrated from the underutilized server to another suitable server in order to turn off the underutilized host.

### 4.2 The Problem formulation

the placement and migration of containers are represented in the cloud system, supposing that there are X hosts in a datacenter, characterized by

$$PM = PM_i(PEs_i, MIPS_i, RAM_i) | i = 1...x \quad (1)$$

and the group of VMs characterized by

$$VM = VM_j(PEs_j, MIPS_j, RAM_j) | j = 1...y \quad (2)$$

and the set of Containers characterized by

$$CNT = CNT_k(PEs_k, MIPS_k, RAM_k) | k = 1...z \quad (3)$$

The current Ram utilization is given by:

$$Ram\ utilization = (total\ Ram - Ram\ used)/total\ Ram \quad (4)$$

Every physical machine (PM) can receive one or several virtual machines (VM), each virtual machine can receive one or several Container(CNTs). The Ram utilization of physical machine $PM_i$ is calculated using Eq 4, and each virtual machine $VM_j$ requires $PEs_j$ processing elements, $MIPS_j$ and MBytes of memory $RAM_j$, each Container $CNT_k$ requires $PEs_k$ processing elements, $MIPS_k$ and MBytes of memory $RAM_k$.

## 5. EXPERIMENTAL SETUP

### 5.1 Containers placement

In this experiment, the configuration consists in using 7 hosts, 25 virtual machines and 75 containers (25 of each type: CONTAINER

-type 1, 2, 3). First, the containers are randomly assigned to the virtual machines which will, in turn, be randomly hosted in hosts servers as shown in the figure 6(a). Several scenarios may be considered.

Table 1. Characteristics of hosts ,VMs and Containers in experiment 1(Placement)

|  | Number | MIPS | PES | RAM |
|---|---|---|---|---|
| Host | 7 | 37274 | 8 | 65536 |
| VM | 25 | 37274 | 2 | 1024 |
| CONTAINER -type 1- | 25 | 4658 | 1 | 128 |
| CONTAINER -type 2- | 25 | 9320 | 1 | 256 |
| CONTAINER -type 3- | 25 | 18636 | 1 | 512 |

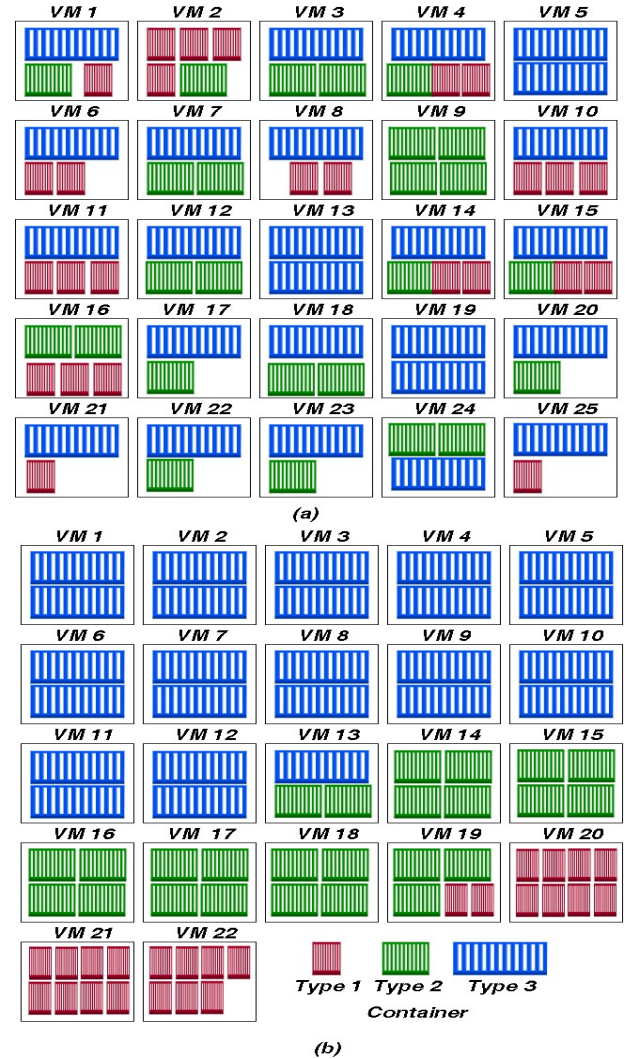

Fig. 6. (a) container allocation randomly.(b) container FFD allocation

Scenario 1: In this scenario, the entire virtual machine RAM memory capacity is used. Container placement can be considered a Bin Packing problem . The bin represents the virtual machine and the items being the containers to be assigned to the Bin. The containers are sorted first in descending order of their Ram memory capacity, and then the containers placement in the VM virtual machine is done using the first fit algorithm . The pseudo-code of containers placement is presented in algorithm 1. the first-fit decreasing algorithm (FFD) is used to obtain an optimal solution. Before applying the FFD algorithm to the allocation problem, the total memory capacity of the containers should not exceed the VM's available memory capacity. This limit is given by:

$$\text{Lower bound} = \sum \text{container(RAM)}/VM(RAM) \quad (5)$$

The Lower bound parameter defines the minimum number of VMs needed to host all the containers. For the proposed configuration, the minimum number of VMs needed to host the 75 containers is as follows:
Lower bound $= 22400/1024 = 21.87 \approx 22$
The First-Fit Decreasing algorithm (FFD) consists in sorting the containers according to their RAM capacities in descending order to assign the containers to the virtual machine while respecting the limit of its memory capacity. the algorithm always start with the first VM if the VM does not contain a place he move to the next VM and so on.
Scenario 2: In this scenario, for issues related to the QoS quality of service, only 90% of the total memory capacity of the virtual machine is used and therefore the limit of the hosting capacity of VM is modified. The limit is given by:
Lower bound $= 22400/1024 * 0.9 = 24.30 \approx 25$
In other words, when using only 90% of the memory capacity of the virtual machine. The number of VMs needed to host the 75 containers goes from 22 to 25 virtual machines (see Figures 6(b) and 7 (b) respectively).
Container placement (FFD). ContainerList ,VmListOutput /: ContainerPlacement container in containerList container.Ram < container.Ram.next Exchange (container, container.next)
max=VmRAM *threshold Virtual Machine in VirtualMachineList container in containerList $VmRamEstimate(Vm, container) <$ max Allocate (Vm ,Container)
Figure 7(a) show that 30 VMs are needed using a threshold of 90 % of their RAM memories to randomly allocate the 75 containers. This number is reduced to only 25 VM when using the FDD algorithm (see Figure 7 (b)). In general, as shown in Figure 9, when the FDD algorithm is used the value of lower bound increases as the percentage of memory capacity of VMs (threshold) decreases. In other words, the number of VMs needed to host all the containers increases as the percentage of the memory capacity of the VM.

### 5.2 VM migrations

Virtual machine migration is a method where VM is moving from one host to another depending on server usage. In order to determine which virtual machine VM is a candidate for migration, a selection strategy VM depending on the threshold is presented. There are two types of VM migration, Virtual Machine Distribution and Virtual Machine Consolidation. For Virtual Machine Distribution: The migration is started when the algorithm detects a hotspot. The migration list is calculated based on the overused hosts, the hosts are sorted by power consumption. The target host is chosen so that there is no overload after migrating the virtual machine to that host. If the source host is still overloaded, the algorithm runs again until the source host is no longer overexploited. Virtual Machine Consolidation: Consolidation algorithm migrates virtual machines from coldspots and turns off the server. Turning off idle servers or under-utilized servers saves energy. Its challenging to reduce number of servers without degrading performance. all the virtual machines of the host should be migrated in order to turn it off [5].

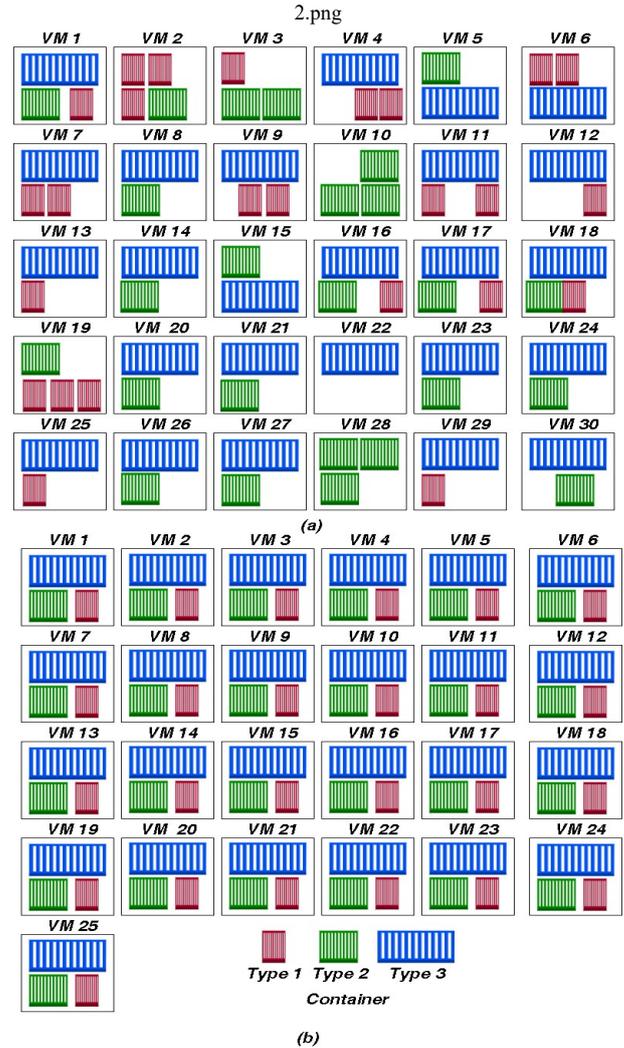

Fig. 7. (a) allocation randomly (threshold 90%) (b) FFD allocation (threshold 90%)

Table 2. Characteristics of hosts , VMs and Containers in experiment 2 (migration).

|  | MIPS | PES | RAM | BW | Max Power |
|---|---|---|---|---|---|
| Host | 10000 | 8 | 9192 | 1000000 | 250W |
| VM | 2000 | 2 | 2048 | 100000 |  |
| CONTAINER | 500 | 1 | 512 | 2500 |  |

In this experiment we concentrate on the Consolidation using VM migration. In this Scenario, Figure 10 (a) shows a configuration consisting of 3 hosts and 7 VMS with a RAM threshold of 90% . Table 2 gives the characteristics used in the proposed configuration. To avoid the waste of host RAM, there is two possibilities: either migrated VMs or containers to optimize memory resources of VMs. Figure 10 (b) shows that the migration of VMs in host 2 did not allow a suitable optimization of memory resources . This is due to the type of isolation of VMs and containers. moreover, migrating VM 6 to Host 1 or Host 2 is not allowed in order to not exceed the threshold which implies the degradation of the quality of service especially the response time.

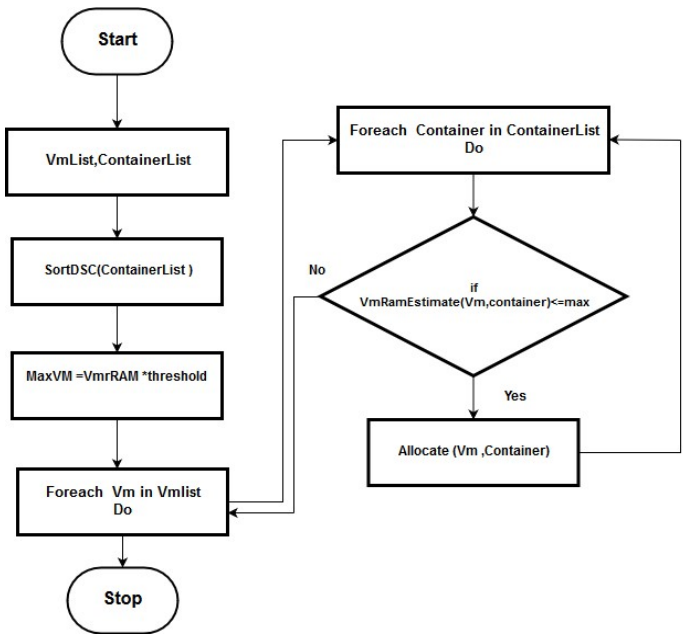

Fig. 8. Flowchart for container placement.

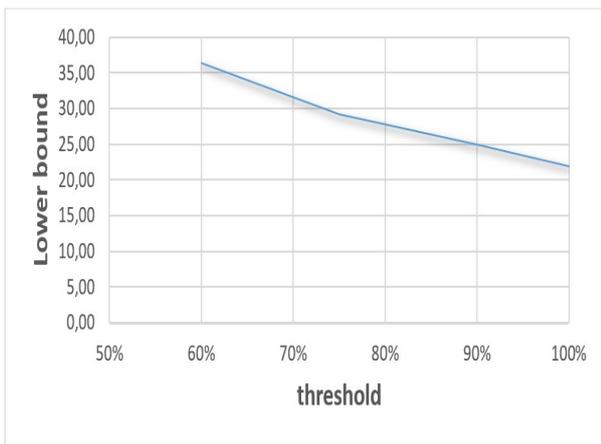

Fig. 9. Lower bound and threshold comparison

### 5.3 Container migrations

in this case the same experiment is repeated using container migration instead of VM migration, in order to move all containers using the container migration. All containers can be moved from host 3 to host1 and host 2 in order to reduce number of servers and consequently reduce the energy consumption without degrading performance (see Figure 11 ). The pseudo-code of containers migration is presented in algorithm 2. As a conclusion,container migration is more optimal when energy consumption is considered.

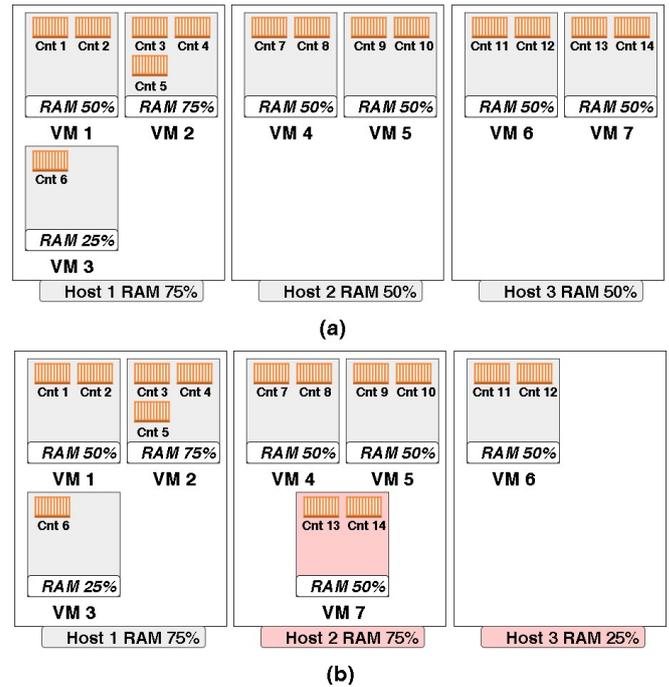

Fig. 10. (a)Illustration of Hosts before Migration (b) Illustration of Hosts after VM Migration

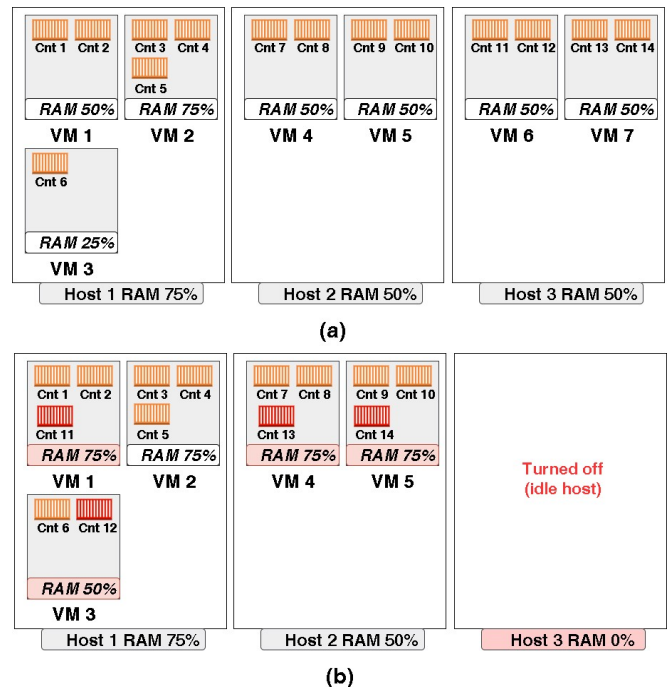

Fig. 11. (a)Illustration of Hosts before Migration (b)Illustration of Hosts after container Migration

Container Migration. Hostlist, VMList,ContainerList Output: MigrationList Host in HostList V M List ← host.getV mList()

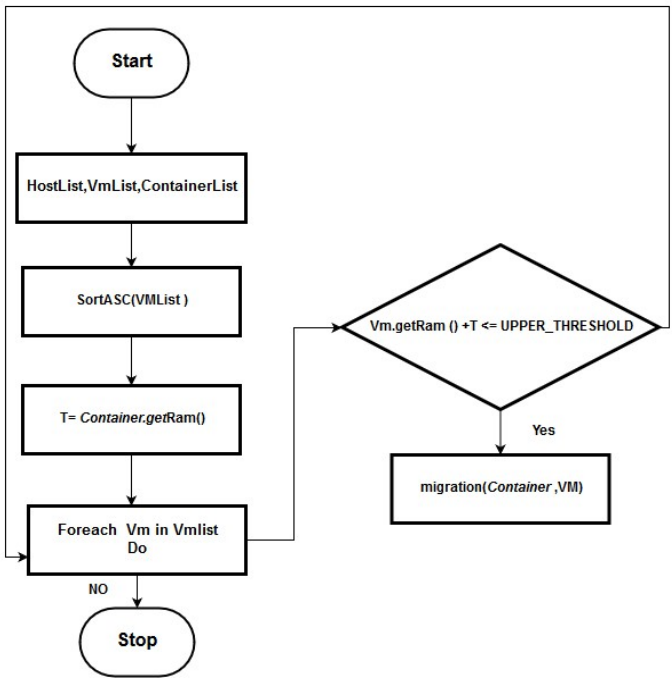

Fig. 12. Flowchart for container migration.

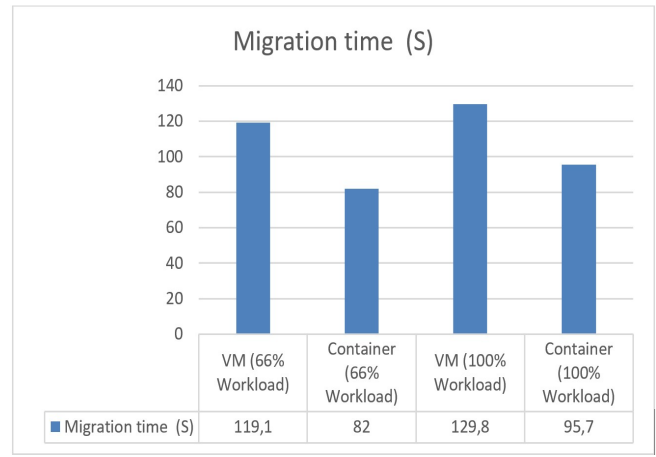

Fig. 13. Comparison of Migration time in VM and Container [18].

VMList.sortAscendingOrderOfRam()  VM in VMList  ContainerList ← VM.getContainerList() T ← Container.getRam()
VM in VMList  V m.getRam() + T  <=UPPERTHRESHOLD
migration(Container ,VM)

### 5.4 Results and discussion

To evaluate the algorithm presented before, simulation results are presented taken away from different approaches, comprise SLA violations, energy consumption, and the number of migrations.

#### 5.4.1 effectiveness comparison parameters.
To measure the efficiency of the algorithms, some metrics are used to evaluate their performance. the following parameters are adopted, to compare the proposed algorithm with others existing algorithm related to migration and placement.

#### 5.4.2 Power consumption and Number of migration.
The first metric is the energy consumption of the physical machine in a data center, it's calculated as follows:

$$P(u) = P\max * Number\ of\ Activated host \quad (6)$$

where Pmax presents 100 % of the power consumption of the server . For the simulations, Pmax is assigned to 250 W, which is considered as normal value. For example, according to the SPEC power benchmark, for the fourth quarter of 2010, the average power consumption at 100 % of use for servers consuming less than 1000 W was around 259 W.
The second metric is the number of migrations started by the manager during the adjustment of the Container placement.
According to the experiences, Container has relatively better performance than VM when total migration time, are considered (see Figure 13 ).
Comparing the two scenarios, the use of container or VM, the migration time is shorter for the container and the total migration time is higher which is very normal, witch can explained by the total number of migration. Regarding energy consumption energy consumed by the datacenter or host is minimal when using the container this is mainly due to the turned off of unused host After migration.(see Figure 14 ).

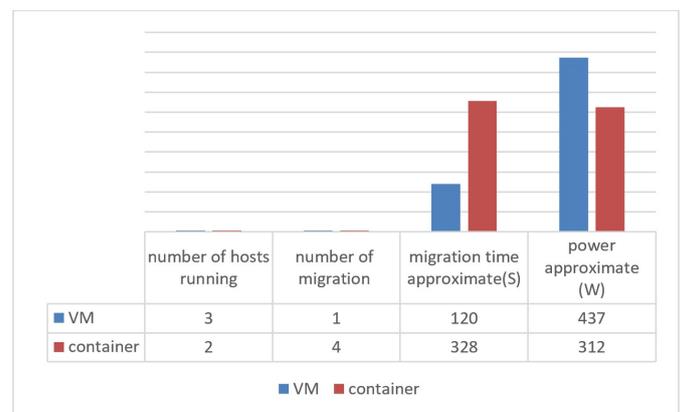

Fig. 14. Comparison of VM and Container Migration.

#### 5.4.3 SLA Level and Downtime.

$$SLA = Total\ uptime - authorized\ Downtime \quad (7)$$

The third metric is SLA Level, a cloud SLA as is mentioned before is an accord between client and cloud service provider that ensures a minimum level of service is maintained. It guarantees levels of reliability, availability, and responsiveness to systems and applications. Uptime is the total time that a service is available and operational. SLA level is defined as assured uptime by the provider , it's considered an important parameter for quality's Measurement of hosting in cloud. According to SLA and Uptime calculator [2], for 99.99 % of SLA level the allowed downtime must not exceed 52 minutes and 36 seconds per year( Daily: 8.6 seconds).
Figure 15 presents the comparison result of the two types of virtualization (Container and VM) based on downtime in the migration process. As the result shows, downtime is higher in VM migration than container migration.

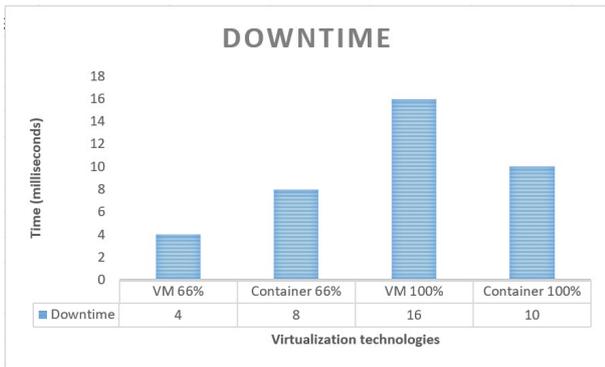

Fig. 15.  Downtime of vm and container[18]

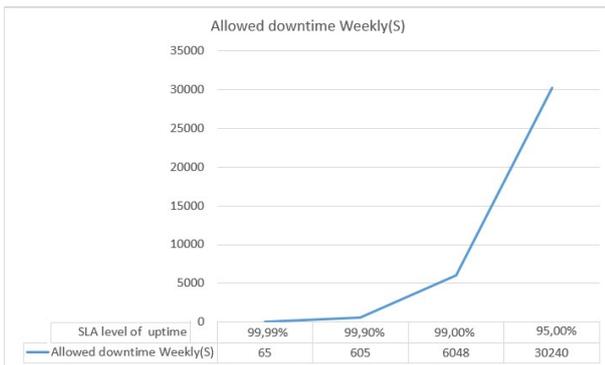

Fig. 16.  SLA level and downtime allowed

according to figure 16, the authorized downtime is limited if the SLA level is high. This is why it recommended to use container migration instead of VM migration to minimize SLA Violation.
From the obtained results, Container has comparatively less downtime than VM. Thereby, Container is more suitable in data centers where demand for availability is a major aspect. In few data centers, even small downtime will impact cloud providers businesses [19]. To meet the requirements of customers , SLA plays a crucial role.

5.4.4   load balancing. Running tasks in containers instead of VMs helps to balance the load of user-sent tasks and also better manages the VM. Migrating containers instead of VMs can also reduce the impact of migration on quality of service by minimizing downtime and also just users of the container on migration can be impacted instead of all VM users.

## 6.  CONCLUSION AND FUTURE WORK

In this work, a system for allocation and migration of containers is presented, taking into account the quality of service requirements of users during the migration of a container from Host or VM to another. In addition,the problem of energy inefficiency due to server underutilization is addressed. The results of the simulation show that the utilization of container migration instead of VMs demonstrates a reduction in energy consumption, and also reduces the migration time which impacts the QoS and reduces the violation of the SLA. The use of the FFD algorithm has allowed us to better optimizer placement of VMs which impact the number of Hosts. The main limitation of the proposed technique is the connection between containers due to the fact that the migration can cause technical problems for the user especially in the case of use of the n-tier application. As a perspective of research work, the tracking of under-utilized Hosts can be done by proposing solutions based on ant colony or genetic algorithms to optimize our architecture based on hybridization between containers and VMS.